# Insight into the structure-property relation of UO$_2$ nanoparticles

Evgeny Gerber[a,b,c], Anna Yu. Romanchuk[c], Stephan Weiss[b], Stephen Bauters[a,b], Bianca Schacherl[d], Tonya Vitova[d], René Hübner[b], Salim Shams Aldin Azzam[b], Dirk Detollenaere[e,f], Dipanjan Banerjee[f,g], Sergei M. Butorin[h], Stepan N. Kalmykov[c] and Kristina O. Kvashnina[a,b,c*]

Highly crystalline UO$_2$ nanoparticles (NPs) with sizes of 2-3 nm were produced by fast chemical deposition of uranium(IV) under reducing conditions at pH 8-11. The particles were then characterized by microscopic and spectroscopic techniques including high-resolution transmission electron microscopy (HRTEM), X-ray diffraction (XRD), high-energy resolution fluorescence detection (HERFD) X-ray absorption spectroscopy at the U M$_4$ edge and extended X-ray absorption fine structure (EXAFS) spectroscopy at the U L$_3$ edge. The results of this investigation show that despite U(IV) being the dominant oxidation state of the freshly prepared UO$_2$ NPs, they oxidize to U$_4$O$_9$ with time and under the X-ray beam, indicating the high reactivity of U(IV) under these conditions. Moreover, it was found that the oxidation process of NPs is accompanied by their growth in size to 6 nm. We highlight here the major differences and similarities of the UO$_2$ NPs properties with PuO$_2$, ThO$_2$ and CeO$_2$ NPs.

## Introduction

Uranium dioxide remains one of the most essential uranium compounds due to its application as a nuclear fuel in most of the commercial nuclear reactors worldwide.[1] Structural chemistry and physics of the U/O system are very complicated but highly important for reactor performance, spent nuclear fuel storage and its further geological disposal. While bulk UO$_2$ has been intensively studied, it is still not clear if the investigated properties persist the same at the nanoscale.[2,3] It is known that actinide (An) nanoparticles (NPs) form aggregates of various sizes.[4,5] In particular, UO$_2$ NPs may be formed by redox reactions from either the reduction of U(VI) by γ- irradiation,[6] minerals,[7–10] microorganisms,[11–17] redox-active chemicals[18] or due to corrosion of metallic U being in contact with water.[4,19,20] It can also be formed via hydrolysis of U(IV) solutions[21–23] or by decomposition of U(IV) compounds.[24]

Under environmental conditions, uranium mineral NPs are found to be ubiquitous and have been identified in a number of studies.[7,25–28] As highly-hydrolysable cation, U(IV) migrates predominantly in the form of pseudo-colloids and intrinsic colloids rather than in the soluble complexed form. UO$_2$ NPs formed as a result of bacteria mediated redox reactions have an influence on U migration in the far-field conditions of repositories. Accidental (like Chernobyl and Fukushima) and routine releases of radionuclides into the environment result in the formation of U oxide NPs.[29–32] It has also been shown that the dissolution of spent nuclear fuel may result in the formation of UO$_2$ NPs that should be taken into account in the performance assessment of repositories,[33] considering that conditions in deep geological repositories are expected to be reducing.

The peculiarities of nanoscale objects affect their properties.[2] Nanoscale UO$_2$ is readily oxidized with the formation of UO$_{2+x}$, while the crystal structure does not significantly alter.[35–37] Similar AnO$_{2+x}$ NPs with a structure close to bulk AnO$_2$ were also observed for plutonium,[38] which is not surprising as both UO$_2$ and PuO$_2$ are isostructural to the fluorite-type fcc structure with a very similar lattice parameter. However, recent publications show that PuO$_2$ NPs do not contain other oxidation states except for Pu(IV)[39,40] and their structural properties are close to the bulk. Similar predictions were made for CeO$_2$ NPs, with the suggestion that CeO$_{2-x}$ NPs were expected to be predominantly composed of Ce(IV), which can be reduced to Ce(III).[41] Later, the absence of the Ce(III) oxidation state was confirmed for NPs even for 2nm particles.[42,43] This could lead to the assumption that there is a similar trend for all highly-hydrolyzed tetravalent Ln or An cations. However, to the best of our knowledge, the tetravalent oxidation state for UO$_2$ NPs has never been proven.

The main difference between U and Pu lies in multivalent behaviour. Under oxidizing conditions, PuO$_2$ is the sole stable oxide, but more than ten stable U binary oxides – UO$_{2+x}$ – are known. Similar to plutonium, CeO$_2$ is the only stable oxide under oxidizing conditions, however both Ce(IV) and Ce(III) ions may present in solution. Later we briefly compare the differences and similarities of various An and Ln oxide NPs properties, based on the results reported here.

[a.] The Rossendorf Beamline at ESRF – The European Synchrotron, CS40220, 38043 Grenoble Cedex 9, France.
[b.] Helmholtz-Zentrum Dresden-Rossendorf (HZDR), Institute of Resource Ecology, PO Box 510119, 01314, Dresden.
[c.] Lomonosov Moscow State University, Department of Chemistry, 119991 Moscow, Russia.
[d.] Institute for Nuclear Waste Disposal (INE), Karlsruhe Institute of Technology, P.O. 3640, D-76021 Karlsruhe, Germany
[e.] Department of Chemistry, X-ray Imaging and Microspectroscopy Research Group, Ghent University, Ghent, Belgium
[f.] Dutch-Belgian Beamline (DUBBLE), European Synchrotron Radiation Facility, 71 Avenue des Martyrs, CS 40220, 38043 Grenoble Cedex 9, France
[g.] Department of Chemistry, KU Leuven, Celestijnenlaan 200F, Box 2404, B-3001 Leuven, Belgium
[h.] Molecular and Condensed Matter Physics, Department of Physics and Astronomy, Uppsala University, P.O. Box 516, Uppsala, Sweden

## Experimental

### Nanoparticle synthesis

The UO$_2$ NPs were synthesized from U(IV) aqueous solution by adding ammonia under reducing conditions. Due to the highly sensitive nature of U(IV) towards oxidation, all synthetic procedures, including the preparation of the samples for the following characterization methods were done in a glovebox under nitrogen atmosphere (<10 ppm O$_2$).

Special care was taken to avoid any contact with oxygen before and during the measurements. U(IV) stock was prepared by galvanostatic reduction of 0.1 M U(VI) in 0.5 M HClO$_4$ (5 hours, 20 mA). The presence of only U(IV) and the stability of the solution were verified by UV-vis spectrometry (AvaSpec-2048x14, Avantes, Fig. S1). Each U(IV) solution (0.1 M and 0.01 M) was divided into two parts. The first set of aliquots was added to 3 M NH$_3$ in the volume ratio 1:10 under continuous stirring. The pH of the 3M ammonia solution was 12.5, but the pH slightly decreased due to the interaction with the U(IV) solution, most likely due to hydrolysis reactions. This set of samples was named "0.1 M/0.01 M U(IV) pH > 11". The second set of aliquots of stock U(IV) was added to water in the volume ratio 1:10, after which several drops of 3 M NH$_3$ were added under continuous stirring to reach pH 8. This set of samples was named "0.1M/0.01M U(IV) pH 8". In all syntheses, mixing rate and vessel geometry were maintained constant. The precipitation process for all samples started shortly (within ten minutes) after addition of all reagents. A black precipitate was formed, and the reaction was continued for about 2 hours to reach equilibrium. Then, the pH and redox potential of the formed suspensions were measured (Table S1). The UO$_2$ reference was made by pressing industrially obtained uranium dioxide powder into a pellet followed by sintering at 1700°C under H$_2$/Ar stream. The industrial uranium dioxide, in its turn, was obtained from UF$_6$ by the gas-flame method, followed by annealing under reducing conditions at 600 – 650 °C.[44] The reference was characterized by X-ray diffraction (XRD) and polarography, the oxygen coefficient of UO$_{2+x}$ was found to be in order of x=0.001.

### Characterization

#### Transmission electron microscopy (TEM)

TEM investigations were performed at the Helmholtz-Zentrum Dresden-Rossendorf (HZDR) using an image-C$_s$-corrected Titan 80-300 microscope (FEI) operated at an accelerating voltage of 300 kV. In particular, selected area electron diffraction (SAED) patterns using a SA aperture of 40 μm and high-resolution TEM (HRTEM) images were recorded.

#### Powder X-ray diffraction measurements.

Powder X-ray diffraction (PXRD) data were collected at room temperature at the HZDR. XRD diffractograms data were collected with a MiniFlex 600 diffractometer (Rigaku, Tokyo, Japan) equipped with a Cu Kα X-ray source (40 keV/15 mA operation for X-ray generation) and the D/teX Ultra 1D silicon strip detector in the Bragg-Brentano θ-2θ geometry at a scanning speed of 0.02 degrees per min. The FWHM and peak position were determined with Fityk software.[45]

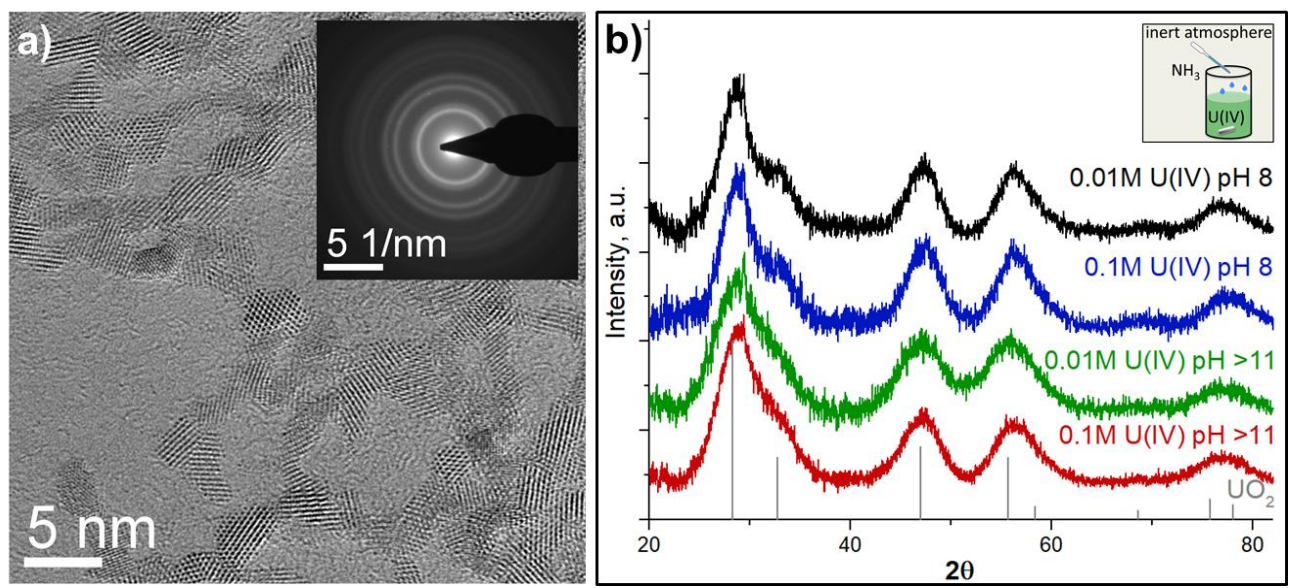

Fig. 1. a) HRTEM image of 0.01M U(IV) pH 8 NPs and corresponding SAED pattern (inset), b) XRD patterns of UO$_2$ reference and the precipitates from U(IV) with different pH and concentrations. The inset shows the schematic drawing of UO$_2$ NPs synthesis.

**X-ray absorption near edge structure (XANES) in high-energy resolution fluorescence detection (HERFD) mode at the U $M_4$ edge and U $L_3$ extended X-ray absorption fine structure (EXAFS) spectroscopy**

The HERFD spectra at the U $M_4$ edge were collected at the CAT-ACT beamline of the KARA (Karlsruhe research accelerator) facility in Karlsruhe, Germany.[46] The incident energy was selected using the [111] reflection from a double Si crystal monochromator. Estimated flux at the sample position was in the order of $10^9$ ph/sec at an incident energy of 3.8 keV.[47] The U HERFD spectra at the $M_4$ edge were obtained by recording at the maximum intensity of the U $M_\beta$ emission line (3339.8 eV) as a function of the incident energy. The emission energy was selected using the [220] reflection of one spherically bent Si crystal analyser (1 m bending radius) aligned at 75° Bragg angle. Samples were prepared and sealed in a special argon-filled container at the licensed laboratory of HZDR and were transported to KARA under inert conditions. All samples were mounted in the form of wet pastes within triple holders with 8 μm Kapton window on the front side, serving as first confinement. Three of such holders were mounted in one larger cell, with 13 μm Kapton window on the front side. (second confinement, Fig. S2) The second confinement chamber was constantly flushed with He. The entire spectrometer environment was contained within a He box to improve signal statistics. An energy range from 3710.5 to 3790.5 eV was scanned with a step size down to 0.1 eV using a 1s dwell time per energy point. All samples were tested for short-term beam damage. First an extended timescan (>2 min with 0.1sec step) above the excitation edge was performed before data collection, to monitor any long-term variations in fluorescence signal. Later a preliminary fast HERFD scan (<2 min) was collected and compared with all HERFD data collected per sample. Based on that procedure, the estimated X-ray exposure time has been derived for each sample.

The U $L_3$ edge (17166 eV) EXAFS spectra were collected at BM26A, the Dutch-Belgium beamline (DUBBLE) at the ESRF (the European Synchrotron) in Grenoble, France.[48] The energy of the X-ray beam was tuned by using a double-crystal monochromator operating in a fixed-exit mode using a Si(111) crystal pair. Measurements were performed in transmission mode with $N_2$/He and Ar/He filled ionization chambers. Energy calibration was performed by recording the EXAFS spectrum of the K-edge of metallic Y (~17038 eV) which was collected simultaneously with the sample scans for each sample. The samples were measured at room temperature using a double-confined, heat-sealed polyethylene holder. Energy calibration, averaging of the individual scans, EXAFS data extraction and fitting were performed with the software package Demeter.[49]

## Results and discussion

Due to chemical reactivity, the ideal structure of $UO_2$ can be easily perturbed (even in the bulk crystal).[35,50,51] The particle size distribution and crystallinity could also differ depending on the synthesis route[6,11,52–54]. The HRTEM data reported in Fig. 1a and Fig S3 confirm that regardless of U(IV) concentration and pH conditions, similar NPs are formed (with respect to their size distribution and crystallinity). A comparison of the SAED patterns (Fig. 1a and Fig. S3 (insets)) and diffractograms from XRD measurements (Fig. 1b) with bulk $UO_2$ show that the crystalline structure of the NPs is similar to that of bulk $UO_2$ (ICDD 03-065-0285). However, the diffraction peaks are broad, indicating the nanosize dimensions of the samples. The crystallite size was estimated from XRD with Scherrer's equation and found to be similar for all samples, varying in the range of 1.7-2.5 nm (Table S2) with respect to the fact that Scherrer's equation is supposed to give information about coherent domains rather than crystallites. Nevertheless, the diffraction peaks of NPs obtained at pH 8 are slightly narrower than those for pH >11, indicating that pH has a small but notable effect on the NP size. It was unexpected as previous research showed that synthesis conditions highly impact the $UO_2$ shape and size. For example, Hu and coauthors[55] made $UO_2$ in the form of NPs, nanoribbons and nanowires, changing the precursor/organic solvent ratio and temperature. There are many other examples, where the size of the obtained $UO_2$ NPs varied from several nm up to several microns depending on the synthesis route (radiolytic reduction, organic precursor-assisted syntheses, U(IV) hydrolysis, biogenic reduction etc.) as well as starting precursors and reaction conditions.[8,13,52–54,56–63] Our HRTEM results (Table S2) also confirm the nanosize of crystallites.

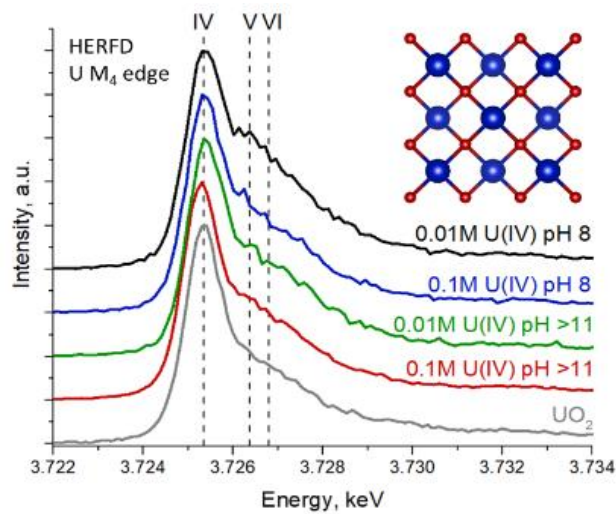

Fig. 2. U $M_4$ HERFD experimental data recorded for the four $UO_2$ NPs samples and compared to a $UO_2$ reference.

For U–O systems, the number of stoichiometric binary oxides and solid solutions with various compositions are known. Uranium upon oxidation may form various oxides with mixed oxidation states of U (like $U_2O_5$, $U_3O_7$, $U_4O_9$, $U_3O_8$).[64,65] The fluorite structure of $UO_2$ can accommodate a large amount of excess oxygen up to $UO_{2.25}$, therefore XRD, giving information about coherent scattering domains, is generally less sensitive to this kind of alterations. In other words, the XRD patterns of $U_4O_9$ and $UO_2$ are very similar and the presence of those species in $UO_2$ NPs cannot be detected by

XRD (Fig. S4). Further oxidation of $UO_{2.25}$ ($U_4O_9$) may lead to the formation of $UO_{2+x}$ oxides with $0.25 < x \leq 0.33$ accompanied by a changing of the crystal structure. Subsequent oxidation may proceed through the formation of $UO_{2.5}$ ($U_2O_5$) and $UO_{2.67}$ ($U_3O_8$) until $UO_3$ is formed.

To further complicate matters, there is a peak broadening effect for NPs in XRD, making reliable analysis nearly impossible, especially in the case of extremely small NPs. That is where synchrotron-based high-energy resolution fluorescence detection (HERFD) X-ray absorption spectroscopy at the U $M_4$ edge really comes in use. Thanks to its high sensitivity, it can easily detect even the tiniest oxidation state impurities, which are present in many uranium oxides ($U_4O_9$, $U_3O_8$, $U_3O_7$).[9,10,46,66–69] The HERFD method at the An $M_4$ edge probes An 3d-5f electronic transitions and is thus highly effective for the detection of the 5f electron configuration and for oxidation state identification.[9,10,39,40,46,66] Moreover, the shapes of the recorded data on various mixed uranium oxides are so distinct[66,67] that recorded HERFD spectra on uranium systems can be straightforwardly analyzed by a fingerprint approach to detect the presence of $U_4O_9/U_3O_8$ impurities in $UO_2$ NPs.

Fig. 2 shows the U $M_4$ edge measurements on four $UO_2$ NP samples compared to the spectrum of the $UO_2$ reference. All spectral features of $UO_2$ NPs are very similar, corresponding to those of the $UO_2$ reference, thus confirming the results from XRD and HRTEM. The shape of the main absorption peak in the HERFD spectrum of $UO_2$ reference shows an asymmetric profile, which was observed before,[66] however for NP spectra the asymmetry of the peak increases, leading to a high-energy shoulder (Fig. S5). It is expected that the asymmetry of the peak originated from partial oxidation and the presence of oxidized uranium species. However, the theoretical calculations of the U(IV) $M_4$ HERFD spectra (c.f. SI, Fig. S6) show that nanoscale distortion or even different coordination environment has a strong correlation with the high-energy shoulder in U $M_4$ HERFD. It is not easy to distinguish between the influence of the presence of higher oxidation state in NPs and the distortion contributions to the asymmetry of the peak. However, theoretical results clearly indicate that the distortion at the surface and random changes of the coordination number for surface atoms will affect the intensity (increase and decrease) of the higher energy U $M_4$ HERFD shoulder. Regardless of the asymmetry origin, one can conclude that U(IV) is the dominant oxidation state for $UO_2$ NPs. To the best of our knowledge, this has never been shown and reported for $UO_2$ NPs at the high sensitivity that HERFD gives for redox speciation.

The crystallinity of the NPs was investigated for structural disorder by uranium $L_3$-edge EXAFS studies (c.f. Fig. S7, Table S3). EXAFS is actively used for uranium[8,10,63,68,70] to investigate the local chemical environment. It has also been used previously to determine the oxidation state of uranium due to the different U(X)-O bond length (where X is the oxidation state of uranium) and static disorder contributions. Our U $L_3$-EXAFS data and shell fit results indicate a $UO_2$-like structure, with characteristic distances 2.33 Å and 3.85 Å for U-O and U-U, respectively (Fig. S7). The absence of other shorter or longer U-O distances suggests that there is no need to invoke different U oxidation states or a substantially different structure (e.g. $U_4O_9$ or U(V)-O). However, the reduced CN for the U-O shell and high Debye-Waller factors are suggestive of particle-size effect and static and thermal disorders, which was previously observed on similar particles.[8,9,13,63,71]

Taking into account: 1) the significant disorder revealed by EXAFS and 2) the theoretical prediction of the distortion effects on the high-energy shoulder of the U $M_4$ HERFD spectra, surface distortion might be the predominant reason for the experimental observation of the intensity variation of the high-energy shoulder in U $M_4$ HERFD data between various $UO_2$ NPs.

**Reactivity of the $UO_2$ NPs**

The reactivity of $UO_2$ NPs under different conditions was studied previously. It was found, that sintering leads to NP growth.[6,13] Rath et al. found that $UO_2$ NPs obtained with γ-irradiation oxidize in several hours under air conditions, while Singer et al. and Wang et al. did not find any changes after 2 months ageing or several days of air exposure.[11,53,60] However, visual observations of the change in residue colour by UV-vis spectroscopy or the U L$_3$ edge XANES (used in previous studies) may not be sufficient to detect other oxidation states of uranium.

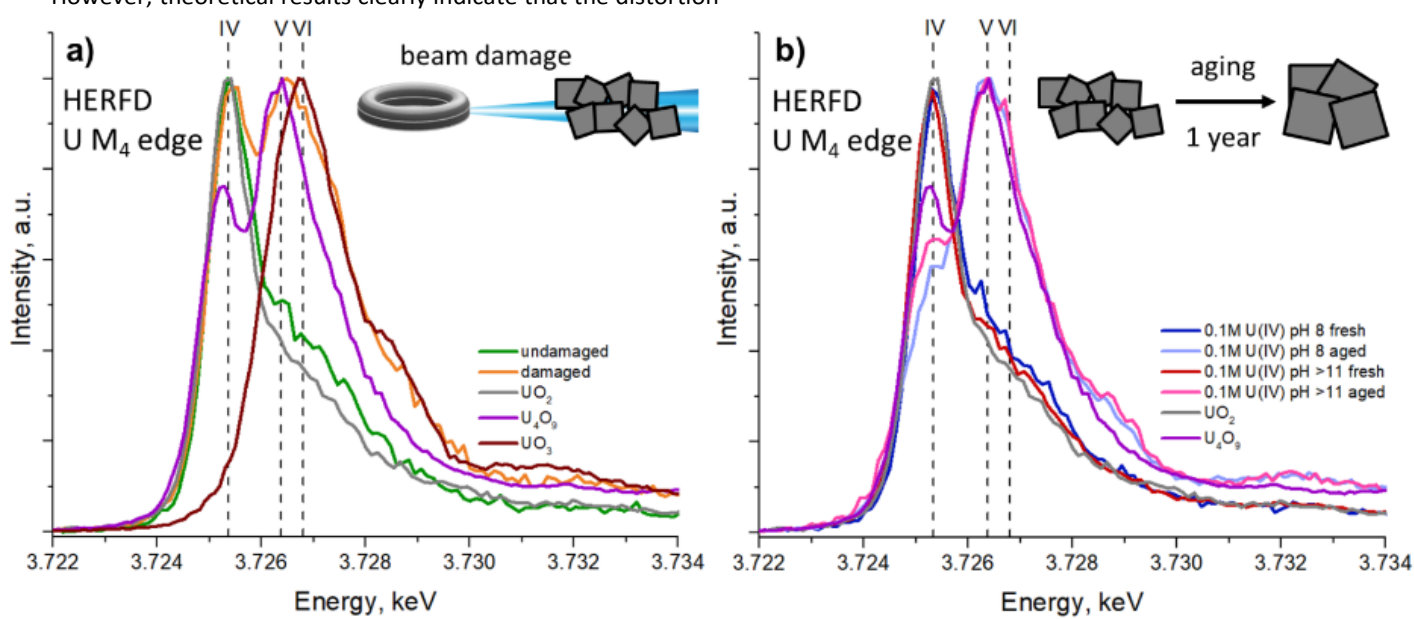

Fig. 3. **HERFD $M_4$ edge spectra:** a) for 0.01M U(IV) pH >11 before and after beam damage with references. b) for fresh and 1-year-aged 0.1M samples with references. The HERFD spectra of the $UO_3$ and $U_4O_9$ references have been reproduced from data, reported by Leinders et al and have shown here for clarity.[60]

Here we investigate in detail the reactivity and phase stability of synthesized UO$_2$ NPs. First, we noticed the impact of the synchrotron X-ray beam on the freshly synthesized materials. In order to verify the damage of samples by X-ray irradiation, all samples were scanned several times at the exact same position to examine how the beam exposure affects the samples. Fig. 3a shows the average of the first U M$_4$ spectra scans on UO$_2$ NPs synthesized from 0.01M U(IV) pH >11 and its comparison with the last scans after 45-60 minutes of X-ray exposure. With each subsequent scan (which takes 5 min) the shoulder of the HERFD high-energy side increases and can even be resolved as an individual component, indicating the partial oxidation of the sample to U(V) and U(VI). Other samples were oxidized under exposure as well, leading to the conclusion that beam exposure is responsible for uranium oxidation in NPs. In order to obtain data on freshly made materials, HERFD measurements have been made on newly synthesized UO$_2$ NPs (reported in Fig. 2), with a short X-ray exposure time (15 min in total over several scans) on samples sealed in a special inert gas-filled container (more info is given in experimental section and SI).

The recorded U M$_4$ HERFD data on oxidized UO$_2$ NPs are reported in Fig.3 and compared to the spectra of UO$_2$, U$_4$O$_9$ and UO$_3$ (reproduced from Leinders et al.[67]). It should be noted that despite the longer duration of L$_3$-edge EXAFS measurements beam damage does not take place in that case due to the lower unfocused beam intensity, compared to M$_4$-edge HERFD measurements. Several scans made on the same sample position were reproducible and oxidation (or significant differences from UO$_2$ structure) was not detected.

Moreover, the stability of NPs over time has also been studied. Two samples synthesized from 0.1M U(IV) concentrations were kept as wet pastes under inert conditions and ambient temperature for a year in closed 2mL plastic tubes with a tiny amount of water left after washing the NPs. Afterwards these samples were analysed with HRTEM, XRD and M$_4$ edge HERFD techniques. It was found that the size of NPs increases (Table S2, Fig. S8-9) after aging (likely, due to the dissolution-precipitation processes[57]), while partial oxidation was observed by HERFD (Fig. 3b). A rise of U(V) contribution in the damaged NPs is shown (Fig. 3a), though it is clear by the difference in peak intensity ratios that the NPs have not fully converted to pure U$_4$O$_9$. Bulk U$_4$O$_9$ has equal amounts of U(IV) and U(V), yet the peak intensities in HERFD are not the same due to the different probability of the absorption process, i.e. different absolute absorption cross sections of U(IV) and U(V).[67] Therefore, a significant amount of U(IV) is still retained in our damaged samples. The HERFD spectra on the aged NPs show higher contribution of the U(V) for compare to U(IV). Overall, it leads to the conclusion that small NPs oxidize to U$_4$O$_9$ and grow up to 6 nm over time. The size of 6 nm was determined by XRD data and is in agreement with HRTEM size estimations (Table S2).

The strong influence of the X-ray beam and the aging behaviour of the UO$_2$ NPs are clear evidences of the low stability of the samples; therefore, special care must be taken to avoid sample oxidation and destructive effects of the X-ray beam. The reasonable solution is to keep samples under reducing conditions as long as possible before performing any experiments, to record relatively quick scans and to choose new sample positions for every scan to limit sample exposure. Measurements under cryogenic conditions might also overcome the issue of beam damage, but this has not been tested yet on the UO$_2$ NPs.

**Comparison between various An and Ln oxide NPs**

Tetravalent cations (Cat) undergo extensive hydrolysis accompanied by the formation of mono- and oligomeric species. Eventually CatO$_2$ NPs originate from aqueous solutions. Besides uranium, the formation of small (2-4 nm) crystalline NPs was observed for cerium,[43] thorium,[72] neptunium[73] and plutonium.[74] Dioxides of these elements demonstrate similar crystallographic properties: fluorite-type structure with similar lattice parameter. However, the redox properties of these elements are different. Thorium is redox inactive, while Ce, U, Np and Pu may be present in different oxidation states. The correlation between UO$_2$ and PuO$_2$ NPs is of high interest as both U(IV) and Pu(IV) are mobile in their colloid forms.[5,75,76] The redox conditions in deep geological repositories are expected to be reducing. Therefore, U(IV), being stable under these conditions, could be a reference for Pu(IV) as well, due to the similarities in An(IV) ionic radii and crystallographic properties of their dioxides. Our investigation shows that there are many resemblances between these An(IV) NPs. It was shown[39] that neither super stoichiometric AnO$_{2+x}$ nor other higher oxide phases are present in PuO$_2$ NPs though it could be expected due to the stability of Pu(IV) under these conditions. Similar behaviour can be predicted for NpO$_2$ NPs, however, to the best of our knowledge the presence of other oxidation states in NpO$_2$ NPs has not been studied yet by HERFD method.

Contrariwise to Pu and especially Np, Ce(III) is stable in aqueous solutions, therefore, one can expect Ce(III) is present in the hydrolysis products. Nevertheless, it was found[42,43], that CeO$_2$ NPs do not contain even slight amounts of Ce(III), leading to the conclusion that CatO$_2$ NPs formed by fast chemical deposition retain Cat(IV) as the dominating oxidation state regardless of their redox affinity. In this study we confirmed, that it is also valid for UO$_2$ NPs, synthesized by fast chemical deposition method under pH 8-11. Those results do not include the possibility that the formation of other phases or other oxidation states takes place under different synthesis conditions.

## Conclusions

It is reasonable to believe that the properties of UO$_2$ in bulk and at the nanoscale are different. Due to a larger surface-to-volume ratio, UO$_2$ NPs are expected to be more reactive and, therefore, to exist as $UO_{2+x}$, with some of the U oxidized at the surface.[35–37] However, it was found that U(IV) is the dominant oxidation state of the $UO_2$ NPs, synthesized by fast chemical deposition method at pH 8-11, but their stability is significantly lower then bulk-$UO_2$ in terms of time and oxidation sensitivity. They are easily oxidized not only in the air but also slowly under inert conditions or during X-ray exposure. Therefore, special care has to be taken while investigating reactions with $UO_2$ NPs and their properties.

The electronic and local structure of the freshly synthesized $UO_2$ NPs in size of 2-3nm was revealed by combination of the U $L_3$-edge EXAFS, U $M_4$-edge HERFD, XRD and HRTEM methods. We show here that the structural and electronic properties of fresh ultra-small $UO_2$ NPs (2-3 nm) are similar to bulk $UO_2$ when inert or reducing conditions restrictions are maintained. It was found that high reactivity of $UO_2$ NPs in time and under X-ray beam exposure leads to the formation of the $U_4O_9$ species complemented by the growth of the NP size to 6 nm. We believe that these findings are beneficial to the fundamental understanding of nuclear fuels and tailoring the functionality of $UO_2$ since most previous studies focus on large-bulk $UO_2$.

## Conflicts of interest

There are no conflicts to declare.

## Acknowledgements

This research was funded by European Commission Council under ERC [grant N759696]. E.G. acknowledges support from RFBR (project number No. 19-33-90127). S.N.K. acknowledges support by the Russian Ministry of Science and Education under grant № 075-15-2019-1891. S.M.B. acknowledges support from the Swedish Research Council (Grant 2017-06465). Authors thank HZDR for the beamtime at CAT-ACT beamline of KARA. Moreover, we acknowledge the help of J. Rothe, A. Beck, J. Galanzew and T. Prüßmann at CAT-ACT beamline of KARA during the HERFD experiment at the U $M_4$ edge. Furthermore, the use of the HZDR Ion Beam Centre TEM facilities is acknowledged.